\documentclass[prb,nofootinbib,twocolumn,superscriptaddress]{revtex4} 

\usepackage{graphicx}
\usepackage{dcolumn}
\usepackage{bm}
\usepackage{threeparttable}
\usepackage{times}
\usepackage{mathptmx}
\usepackage{lscape}
\usepackage{natbib}
\usepackage{amsmath}
\usepackage{amssymb}
\usepackage{braket}
\usepackage{comment}
\usepackage{color}

\def\degree{\kern-.2em\r{}\kern-.3em}

\begin{document}

\title{ Unavoidable Canonical Nonlinearity Induced by Gaussian Measures Discretization  }

\author{Koretaka Yuge}
\affiliation{
Department of Materials Science and Engineering,  Kyoto University, Sakyo, Kyoto 606-8501, Japan\\
}%

\begin{abstract}
{ 
When we consider canonical averages for classical discrete systems, typically referred to as substitutional alloys, the map $\phi$ from many-body interatomic interactions to thermodynamic equilibrium configurations generally exhibits complicated nonlinearity. 
This canonical nonlinearity is fundamentally rooted in deviations of the \textit{discrete} configurational density of states (CDOS) from \textit{continuous} Gaussian families, and has conventionally been characterized by the Kullback-Leibler (KL) divergence on discrete  statistical manifold. Thus, the previous works inevitably missed intrinsic nonlinearities induced by discretization of Gaussian families, which remains invisible within conventional information-geometric descriptions.
In the present work, we identify and quantify such unavoidable canonical nonlinearity by considering a transport cost induced by the specific local-map (LM) discretization scheme, based on the 2-Wasserstein framework with a cost function aligned with the Fisher metric for Gaussian families. In the limit of vanishing discretization scale $d\to 0$, we derive an explicit expression for this specific transport cost: $\widetilde{W}_{2}=d\sqrt{\textrm{Tr}\left( \Gamma^{-1} \right)/12 }$, where $\Gamma$ denotes covariance matrix of the Gaussian. We show that this limiting transport cost admits a clear geometric interpretation on the statistical manifold, corresponding to a KL divergence associated with the expected parallel translations of continuous Gaussian. In addition, we confirm that this $\widetilde{W}_{2}$-KL correspondence admits a natural generalization beyond Gaussian families, provided that the cost function is aligned with the Fisher metric of an underlying statistical submanifold and the discretization scale links to infinitesimal parameter variations. The correspondence demonstrates that the geometric distortion of the local measure induced by discretization---while extrinsic to information geometry alone---can be naturally characterized by a standard KL divergence.}
\end{abstract}

\maketitle

\section{Introduction}

Canonical averages play a central role in statistical thermodynamics, providing a fundamental link between microscopic interactions and thermodynamic equilibrium configurations.
For classical discrete systems with $f$ structural degrees of freedom (SDFs) on a given lattice, such as substitutional alloys, this correspondence is expressed as
\begin{eqnarray}
\label{eq:can}
\Braket{ q_{p}}_{Z} = Z^{-1} \sum_{i} q_{p}^{\left(i\right)} \exp \left( -\beta U^{\left(i\right)} \right),
\end{eqnarray}
where $\{ q_{1}, \cdots, q_{f} \}$ denotes a complete set of structural coordinates, $\Braket{\cdot}_{Z}$ the canonical average, $\beta$ the inverse temperature, and
$Z=\sum_{i}\exp\left(-\beta U^{\left(i\right)}\right)$ the partition function, with the summation taken over all microscopic configurations $i$.

When the structural coordinates are chosen as a complete orthonormal basis, such as those of the generalized Ising model (GIM),\cite{ce} they correspond to a  complete set of symmetry-independent cluster correlation functions defined on the underlying lattice. The potential energy of configuration $k$ is then expressed as
\begin{eqnarray}
\label{eq:u}
U^{\left(k\right)} = \sum_{j=1}^{f} C_{j} q_{j}^{\left(k\right)},
\end{eqnarray}
where the expansion coefficients are given by inner products $C_{j}=\Braket{U|q_{j}} = \sum_{l}\zeta^{-1} U^{\left( l \right)}\cdot q_{j}^{\left( l \right)}$, i.e., summation over all microscopic configurations $l$ with normalization constant $\zeta$ defining the inner product. 
Therefore, under the GIM description, $f$ denotes the number of independent GIM basis functions.

Introducing the vectors ${Q}_{Z}=(\Braket{ q_{1}}_{Z},\cdots,\Braket{ q_{f}}_{Z})$ and
${U}=(C_{1},\cdots,C_{f})$, the canonical average in Eq.~(\ref{eq:can}) defines a map
\begin{eqnarray}
\label{eq:map}
\phi : {U} \mapsto {Q}_{Z}.
\end{eqnarray}
It is well known that this map is generally nonlinear.
The behavior of this map is governed by the configurational density of states (CDOS), which is the discrete distribution of all possible microscopic configurations in the configurational coordinate space
$\{q_1,\cdots,q_f\}$, independent of temperature and interatomic interactions.
Hereafter, we denote the structural-coordinate vector by
$q=(q_1,\cdots,q_f)$.
Only in the hypothetical cases where the CDOS is represented by a continuous multivariate Gaussian distribution,
$\phi$ reduces to a globally linear map.\cite{ig} Note that the Gaussian distribution does not represent the
actual CDOS of a discrete system, but rather serves as a hypothetical continuous reference distribution to measure the nonlinearity.
In realistic discrete systems, the CDOS inevitably deviates from the Gaussian reference distribution due to the discrete nature of configuration space imposed by lattice constraints.

Existing computational approaches combining first-principles calculations and statistical thermodynamics enable accurate prediction of equilibrium properties for complex alloy systems. Such approaches typically employ the GIM together with optimization techniques such as cross-validation, genetic algorithms, and machine-learning regression.\cite{cm1,cm2,cm3,cm4,cm5,cm6}  However, they do not by themselves clarify the origin or structure of the nonlinearity inherent in the canonical map $\phi$. This is particularly true from the perspective of configurational geometry (CG), which is informed by the CDOS on a given lattice. To address this problem, we previously introduced the concept of canonical nonlinearity (CN),\cite{ig,igt} which quantifies the nonlinear character of the canonical map based on information geometry, in particular the Kullback-Leibler (KL) divergence:\cite{klo} The  CN measure can be evaluated solely from the CDOS landscape, without requiring any information about interatomic interactions or temperature. 
This viewpoint is important because the CDOS, and hence the CG underlying the canonical map $\phi$, is determined solely by the lattice and composition. Thus, studying $\phi$ from the viewpoint of CG offers a route to systematically  understanding features of thermodynamic equilibrium states \textit{a priori}, based only on configurational information of the system, without specifying temperature or interaction parameters.
Indeed, our previous studies have demonstrated that this viewpoint can provide systematic prediction of short-range ordering (SRO) in alloys using special atomic configurations that are independently determined from temperature and interactions, \cite{em2, ps1} as well as for organizing SRO tendencies in multicomponent alloys using an extended, CG-based definition of the atomic radius ratio.\cite{ratio} We have also shown that the nonlinearity of $\phi$ itself contains information that cannot be explained solely by the number of configurations associated with the number of constituent elements for systems with $R\ge 3$ components, as quantified by the KL divergence.\cite{ig}

In this framework of the CN, the nonlinearity is evaluated by comparing the discrete CDOS of a real system with a reference Gaussian distribution having the same mean and covariance matrix, where the Gaussian is discretized on the same configurational support as the real CDOS.\cite{ig} This information-theoretic formulation has also enabled thermodynamic interpretations of canonical nonlinearity in our previous study.\cite{igt} However, while this framework successfully captures deviations of the CDOS from Gaussian families, it necessarily includes additional contributions arising from the discretization of continuous Gaussian families themselves.

The objective of the present study is not to redefine canonical nonlinearity (CN) itself, but to isolate and quantify the unavoidable contribution originating solely from the discretization process of Gaussian families. Since the existing formulation of CN is based on the KL divergence, it is essential to represent this contribution consistently within the same information-theoretic framework. The difficulty, however, is that discretization is not an intrinsic operation within information theory, but an external transformation imposed on the sample space. Consequently, the discretization-induced contribution cannot be evaluated directly by information-theoretic quantities alone. The central question addressed in the present study is therefore whether such an external geometric contribution can be incorporated into the existing CN framework without introducing arbitrary choices in the quantification of the discretization process.


In the present work, we define the discretization-induced geometric contribution for Gaussian families as an unavoidable canonical nonlinearity (UCN). To quantify this contribution, we first evaluate the geometric cost associated with discretization based on the framework of optimal transport. 
Here, rather than evaluating the exact 2-Wasserstein distance $W_{2}$ obtained by minimizing over all admissible transport plans, we focus on the specific transport cost $\widetilde{W}_2$ induced by a local-map discretization scheme, as detailed below. 
We then show that, under a transport cost determined in a non-arbitrary manner within the present framework, this geometric quantity admits a consistent local information-geometric representation of $\widetilde{W}_{2}$--KL correspondence. This representation would enable the unavoidable discretization-induced contribution to be incorporated consistently into the existing information-theoretic framework of CN.

It should be noted that the relationship between optimal transport and information geometry, including connections between Wasserstein distances and KL divergence, has been actively investigated in previous studies. In particular, Wasserstein gradient flows, geometric formulations on spaces of probability distributions, and variational approaches have established deep connections between optimal transport and information geometry.\cite{otto,ags,jko} However, these studies primarily focus on the intrinsic geometry of probability distributions, where probability measures themselves are treated as geometric objects and their mutual relations are characterized through distances or transformations. The present work addresses a different problem: the geometric contribution induced by an external transformation of representation, namely the projection of a continuous distribution onto a discrete support. Although such a transformation can itself be quantified by optimal transport, the resulting transport cost is an extrinsic quantity associated with a transformation of the sample space. The central issue addressed here is therefore not the quantification of a distance between probability measures alone, but how such an externally induced geometric contribution can be consistently identified with an intrinsic information-geometric quantity without introducing additional arbitrary structures.

In the following part of this paper, we also show that the derived $\widetilde{W}_{2}$-KL correspondence can be extended beyond Gaussian families, under the condition that the cost function is aligned with the Fisher metric. 
We will see that while the present mathematical derivations are elementary, the underlying conceptual structure relies on nontrivial identifications; readers primarily interested in the interpretation may wish to consult the Conceptual Position in the final part.

\section{Derivation and Concept}
\subsubsection*{Unavoidable Canonical Nonlinearity}

Here, we briefly clarify the concept of canonical nonlinearity and its unavoidable contribution, referred to as the unavoidable canonical nonlinearity (UCN). 
Let  $P\left( q \right)$ denote the CDOS of a realistic discrete system, given as a discrete probability distribution
We also introduce a reference continuous Gaussian distribution $P_{c}\left( q \right)$, which shares the same mean vector and covariance matrix $\Gamma$ as the realistic CDOS.
It is well known that when the CDOS is exactly given by the continuous Gaussian $P_{c}$, the canonical average map $\phi$ in Eq.~\eqref{eq:map} becomes\cite{em2}
\begin{eqnarray}
^{\forall}{ C_{j} }, \ {Q}_{Z} = \left( -\beta \Gamma \right) \cdot {U},
\end{eqnarray}
which implies that $\phi$ reduces to a globally linear map.
From this viewpoint, canonical nonlinearity originates from deviations of the realistic CDOS from Gaussian families.

Based on this observation, previous studies have quantified canonical nonlinearity by measuring the difference between 
$P$ and $P_{c}$ using the KL divergence. 
Since the KL divergence requires the two distributions to be defined on the same support (or to satisfy an inclusion relationship of supports), the continuous Gaussian $P_{c}$ has been discretized on the same configurational support as $P$, yielding a discretized Gaussian distribution $P_{d}$. 
Accordingly, canonical nonlinearity has been characterized by the KL divergence $D\left( P:P_{d} \right)$, which subsequently extends to the KL divergence between the canonical distributions induced by the CDOSs $P$ and $P_{d}$.

While this framework successfully captures non-Gaussian features inherent to the realistic CDOS, it inevitably includes an additional contribution originating from the discretization of the continuous Gaussian distribution itself.
This contribution cannot be attributed to intrinsic non-Gaussian feature of the CDOS.
Instead, it reflects  geometric distortion of local measure for continuous Gaussian induced by the discretization process.

We identify this contribution as the unavoidable canonical nonlinearity, UCN.
By construction, the UCN arises solely from discretizing continuous Gaussian families onto a discrete configurational support.
Consequently, this contribution is fundamentally inaccessible to conventional information-geometric approaches formulated on discrete statistical manifolds, which compare probability weights only on fixed discrete supports.
To isolate and quantify the UCN, it is therefore necessary to compare a continuous Gaussian distribution with its discretized counterpart in a framework that satisfies the following requirements:
(i) it captures not only probability weights but also the geometric rearrangement of probability mass induced by discretization, and
(ii) it remains consistent with, or admits a reinterpretation within, the existing information-geometric description on statistical manifolds.
These considerations naturally motivate the present use of optimal transport theory.

As will be shown, the transport cost adopted in the present framework is constrained by the correspondence between the transport geometry and the information geometry associated with infinitesimal indistinguishability. The resulting transport cost is then mapped into its local information-geometric representation, establishing the $\widetilde{W}_{2}$--KL correspondence derived in the following.

\subsubsection*{Fisher-Aligned Transport Cost on the $W_{2}$ Framework}

Optimal transport theory provides a natural framework for comparing probability distributions defined on different supports, such as continuous and discrete distributions.
To quantify the unavoidable geometric contribution introduced above, we first employ the 2-Wasserstein distance $W_{2}$.
The conceptual basis of the present transport formulation, including the transport cost adopted below, will be discussed in the final section ("Conceptual Position").

Let $Q$ and $R$ be probability measures on $\mathbb{R}^{f}$.
Then the squared 2-Wasserstein distance is given by\cite{w2}
\begin{eqnarray}
W_{2}^{2}\left( Q,R \right) = \inf_{\pi \in \Pi\left( Q,R \right)} \int c\left( x,y \right)\, d\pi\left( x,y \right),
\end{eqnarray}
where $\Pi(Q,R)$ denotes the set of all couplings (or transport plan) with marginals $Q$ and $R$, and $c\left( x,y \right)$ denotes cost  function, i.e., transport cost from $Q\left( x \right)$ to $R\left( y \right)$. 
To ensure the compatibility with statistical manifold, we introduce a cost function as the quadratic form weighted by the inverse covariance matrix $\Gamma^{-1}$ of Gaussian family, namely,
\begin{eqnarray}
\label{eq:cost}
c\left( x,y \right) = \left( x-y \right)^{\mathrm{T}} \Gamma^{-1} \left( x-y \right).
\end{eqnarray}
This choice can be viewed as replacing the standard Euclidean squared distance $\left|x-y\right|^2$ in the definition of the conventional 2-Wasserstein distance by the squared distance induced by the Fisher metric on the Gaussian statistical manifold. Note here that the $W_{2}$ distance between two Gaussian of $g\left( \mu_{1},\Gamma \right)$ and $g\left( \mu_{2},\Gamma \right)$ under the Euclidean metric is known as $\left|\mu_{1}-\mu_{2}\right|^{2}$.\cite{we}
 When we consider Gaussian families, its landscape is completely determined by mean vector $\mu$ and covariance matrix $\Gamma$. Considering that (i) $\mu\in\mathbb{R}^{f}$ and $\Gamma\in\mathbf{M}_{f,f}\left( \mathbb{R} \right)$ and (ii) transport object is in $\mathbb{R}^{f}$, 
the relevant Fisher metric is associated with translations of the mean while fixing covariance matrix, which reduces to the constant metric tensor $\Gamma^{-1}$.\cite{aig}

To isolate the unavoidable contribution arising solely from discretization,
we introduce an auxiliary geometric construction that maps the continuous Gaussian reference distribution onto the discrete support of the realistic CDOS.
This construction is introduced solely for quantifying the discretization-induced contribution and is not intended to describe the physical origin or generation process of the CDOS itself.
Under the auxiliary construction below, we consider the 2-Wasserstein distance
$W_{2}(P_{c},P_{d})$ between a continuous multivariate Gaussian distribution $P_{c}\left(q\right)$ with mean vector $\mu$ and covariance matrix $\Gamma$, and its discretized counterpart $P_{d}\left(q'\right)$.
Here, again, $P_{c}$ is projected onto the same discrete support as the realistic CDOS.
Here, for simplicity, this discrete support is represented by partitioning the configuration space (i.e., correlation space)
$\simeq\mathbb{R}^{f}$ into hypercubic cells $V_k$ of side length $d$.
The discretization scale $d$ corresponds to the spacing between neighboring accessible values of the correlation functions in the discrete support of the CDOS.
Note that the spacing of the discrete support generally differ among different correlation directions on the underlying lattice.
Accordingly, the corresponding discretization cells are generally hyperrectangular with side lengths $d_i$.
The hypercubic representation corresponds to the isotropic case, $d_i=d$.
Then, the present auxiliary construction employs the discretized distribution defined by
\begin{eqnarray}
\label{eq:disc}
^{\forall}k,\;
P_d(q'_k)
=
\int_{V_k}P_c(q)\,dq,
\end{eqnarray}
where $q'_k$ denotes the representative point of the cell $V_k$.

We subsequently consider the discretization of continuous distribution $P_{c}\left( q \right)$ into the discretized counterpart $P_{d}\left( q' \right)$ at vanishing discretization limit $d\to0$.
Starting from the $W_{2}$ framework defined above, we consider the transport cost induced by the present local-map (LM) discretization scheme of Eq.~\eqref{eq:disc}, which we denote by $\widetilde{W}_{2}$.
Since each discretization cell $V_{k}$ is represented by its representative point $q'_{k}$, the LM induces a local projection from every point within the cell onto that representative. The corresponding squared transport cost is therefore given by
\begin{eqnarray}
\label{eq:w-int}
\widetilde{W}_{2}^{2}\left( P_{c}, P_{d} \right) = \sum_{k} \int_{V_{k}} \left( q-q'_{k} \right)^{\textrm{T}} \Gamma^{-1} \left( q-q'_{k} \right) P_{c}\left( q \right) dq.
\end{eqnarray}
Here, $\widetilde{W}_2$ refers specifically to the cost associated with the LM transport map and should be distinguished from the exact 2-Wasserstein distance $W_{2}$ obtained by minimizing the transport cost over all admissible transport plans. No optimization over admissible transport plans is performed in defining $\widetilde{W}_2$.
When we introduce the following variable transform
\begin{eqnarray}
u = q-q'_{k},\ ^{\forall}i: \ \left|u_{i}\right| \le \frac{d}{2}
\end{eqnarray}
with representative hypercubic as $V_{0}$, and taking Taylor series expansion of $P_{c}\left( q'_{k} + u \right)$ around $q'_{k}$, we can reasonably retain the leading order in $d$, thereby 
\begin{eqnarray}
\label{eq:w2}
\widetilde{W}_{2}^{2}\left( P_{c}, P_{d} \right) &=& \sum_{k} P_{c}\left( q'_{k} \right) \int_{V_{0}} u^{\textrm{T}}\Gamma^{-1} u du \nonumber \\
&=& \left\{ \sum_{k} P_{c}\left( q'_{k} \right) \right\} \left\{ \sum_{i}\left( \Gamma^{-1} \right)_{ii} \frac{d^{f+2}}{12} \right\} \nonumber \\
&=& d^{2}\frac{1}{12}\textrm{Tr}\left( \Gamma^{-1} \right).
\end{eqnarray}
The last equation can be obtained since at $d\to 0$,
\begin{eqnarray}
\sum_{k} P_{c}\left( q'_{k} \right)d^{f} = \int_{\mathbb{R}^{f}} P_{c}\left( q \right) dq =1.
\end{eqnarray}
A brief numerical check of the derived expression of Eq.~\eqref{eq:w2} against the original integral form of Eq.~\eqref{eq:w-int} is provided in Appendix A.

Notably, this expression is universal: it depends only on the covariance matrix of the Gaussian distribution and the discretization scale $d$. 
This result provides a quantitative measure of the UCN induced purely by discretization.

\subsubsection*{Information-Geometric Interpretation of $\widetilde{W}_{2}$}
We then provide an information-geometric interpretation of the derived transport cost in Eq.~\eqref{eq:w2}, clarifying its correspondence to the KL divergence associated with \textit{expected} parallel translations of Gaussian distributions.
Under fixed $\Gamma$, the KL divergence between the following two Gaussian distribution is exactly given by\cite{kl}
\begin{eqnarray}
^{\forall}\delta\mu\in \mathbb{R}^{f}, \ D\left( P_{c}\left( \mu+\delta\mu,\Gamma \right) : P_{c}\left( \mu, \Gamma \right) \right) = \frac{1}{2}\delta \mu^{\textrm{T}} \Gamma^{-1} \delta \mu. \nonumber \\
\quad
\end{eqnarray}
This relation certainly shows that under fixed $\Gamma$, the KL divergence provides a well-known quadratic form with the Fisher metric tensor $\Gamma^{-1}$ as discussed above.  To derive the transport cost $\widetilde{W}_{2}$ in Eq.~\eqref{eq:w2}, we perform discretization of Gaussian for $f$-dimensional hypercubic with side length $d$. At the side of statistical manifold, we first extend the translation magnitude of Gaussian $\delta\mu$ to an i.i.d. random vector with probability density $\rho$ given by the uniform distribution, namely,
\begin{eqnarray}
\label{eq:rho}
\rho\left( \delta\mu \right) = \textrm{Unif}\left[ -\frac{d}{2}, \frac{d}{2} \right]^{f} \ \left( \textrm{i.i.d.} \right).
\end{eqnarray}
Under this extended definition, we consider the following expectation:
\begin{eqnarray}
\mathbb{E}_{\rho} \left[ \delta\mu^{\textrm{T}}\Gamma^{-1}\delta\mu \right]  = \sum_{i,k}\left( \Gamma^{-1} \right)_{ik} \mathbb{E}_{\rho}\left[ \delta\mu_{i}\delta\mu_{k} \right]. 
\end{eqnarray}
Since $\mathbb{E}_{\rho}\left[ \delta\mu_{i}\delta\mu_{k} \right]$ corresponds to $\left( i,k \right)$-component of the covariance matrix for $\rho\left( \delta\mu \right)$, from Eq.~\eqref{eq:rho}, we can write:
\begin{eqnarray}
&&\mathbb{E}_{\rho}\left[ \delta\mu_{i}\delta\mu_{k} \right]_{i\neq k} = 0\nonumber \\
&&\mathbb{E}_{\rho}\left[ \delta\mu_{i}\delta\mu_{i} \right] = \int_{-d/2}^{d/2} \frac{1}{d} x^{2} dx = \frac{1}{12}d^{2},
\end{eqnarray}
thereby 
\begin{eqnarray}
\mathbb{E}_{\rho} \left[ \delta\mu^{\textrm{T}}\Gamma^{-1}\delta\mu \right] = d^{2}\frac{1}{12}\textrm{Tr}\left( \Gamma^{-1} \right).
\end{eqnarray}
We therefore obtain the important relationships at $d\to 0$:
\begin{eqnarray}
\label{eq:rel}
\widetilde{W}_{2}^{2}\left( P_{c}\left(\mu,\Gamma\right), P_{d} \right) &=& 2\cdot \mathbb{E}_{\rho} \left[ D\left( P_{c}\left( \mu+\delta\mu,\Gamma \right) : P_{c}\left( \mu, \Gamma \right) \right) \right] \nonumber \\
&=&d^{2}\frac{1}{12}\textrm{Tr}\left( \Gamma^{-1} \right). 
\quad
\end{eqnarray}
Eq.~\eqref{eq:rel} certainly exhibits that the squared transport cost between the continuous Gaussian distribution and its discretized counterpart corresponds to twice the KL divergence averaged over continuous Gaussian translations generated by the discretization procedure.
This result establishes a clear geometric interpretation of the UCN: The cost to project continuous Gaussian onto a discrete configurational support is given by the cumulative information-geometric distance for expected parallel translations within the Gaussian statistical manifold. These contributions are invisible to existing KL-based comparisons restricted to discrete statistical manifolds, but are naturally captured by the present transport-information-geometric framework.

In a realistic discrete system on a lattice, the following three features are unavoidable:
(i) $d$ takes a nonzero positive value, (ii) the domain of the CDOS is bounded, and (iii) the domain is generally asymmetric.
All of these features are dominated by the underlying lattice.
Therefore, to address the effect of UCN in realistic systems, numerical approaches,
such as systematic comparisons between Eq.~\eqref{eq:rel} and Eq.~\eqref{eq:w-int} under various conditions,
are essential for addressing the individual contributions of these three lattice-induced effects.

The relation between the UCN contribution and the canonical nonlinearity considered in previous studies should also be clarified here: The UCN contribution and the canonical nonlinearity previously considered in discrete systems characterize different aspects of the canonical nonlinearity, despite both being formulated in terms of the KL divergence. Whereas the former originates from a representation change from continuous Gaussian to discretized counterpart, the latter measures the discrepancy between the actual CDOS and its Gaussian reference on the same discrete support. A direct quantitative comparison between these two contributions is therefore not yet established. Clarifying whether and how they combine, including possible additive or non-additive relations, is an important subject for our future work.

\subsubsection*{Generalization of the $\widetilde{W}_{2}$-KL Correspondence beyond Gaussian Families}
In the previous sections, we demonstrated that the discrepancy between a continuous Gaussian distribution and its discretized counterpart, measured by the LM-induced transport cost with a cost function aligned with the Fisher metric, converges in the vanishing discretization limit $d \to 0$ to the expectation of the KL divergence between infinitesimally translated continuous Gaussians.
Although these derivations were presented for Gaussian families, the underlying mathematical structure suggests a broader applicability.
We here briefly discuss the conditions under which the $\widetilde{W}_{2}$--KL correspondence can be generalized.

We consider the problem of quantifying the difference between a continuous distribution $P_{\xi}\left(q\right)$ and its discretized counterpart $P'\left(q'\right)$, based on the $\widetilde{W}_{2}$ and the KL divergence, following the strategy developed for Gaussian families.
We introduce the quadratic cost function for the 2-Wasserstein distance $W_{2}(P_{\xi}, P')$ as
\begin{eqnarray}
\label{eq:qq}
c\left(q,q'\right)= \left(q-q'\right)^{\mathrm{T}} \Omega \left(q-q'\right).
\end{eqnarray}
To define the cost function in the form of Eq.~\eqref{eq:qq}, the Fisher metric $\Omega$ is constructed with respect to parameters $\xi = \left(\xi_{1}, \cdots, \xi_{f}\right)$ that can be directly related to the coordinates of the underlying support, where $f$ denotes the dimension of the support.
More precisely, we consider parametrizations that transform covariantly under invertible linear transformations of the support, i.e.,
\begin{eqnarray}
^{\forall}A \in \textrm{GL}\left(f\right): \quad q \mapsto A q \quad \Rightarrow \quad \xi \mapsto A \xi,
\end{eqnarray}
where $\textrm{GL}\left( f \right)$ denotes the general linear group consisting of all invertible $f\times f$ matrices.
Such parametrizations include translational and expectation parameters, which belong to the same vector representation as the support coordinates.
This condition ensures that the metric $\Omega$ captures the sensitivity of the distribution consistently with respect to displacements in the sample space.
We assume that, in these coordinates, the corresponding $f \times f$ Fisher metric $\Omega$ is positive-definite and varies smoothly under infinitesimal changes in $\xi$.

Then, we discretize the underlying continuous space for $P_{\xi}$, by using $f$-dimensional hypercubic with side length $d$, and consider 
its limit of $d\to 0$. 
Under this discretization, we introduce the following linear relationship between $\delta\xi$ and $d$:
\begin{eqnarray}
^{\forall}k: \ \delta\xi_{k} =  a\cdot d,
\end{eqnarray}
where $a>0$ is a constant whose specific value is irrelevant for the following analysis. 
Then we extend $\delta\xi$  to an i.i.d. random vector with its distribution $\rho$, for example, 
\begin{eqnarray}
\rho\left(\delta\xi\right) = \textrm{Unif} \left[ -\frac{a\cdot d}{2}, \frac{a\cdot d}{2}\right]^{f} \ \left(\textrm{i.i.d.}\right),
\end{eqnarray} 
where the specific form of the distribution is not essential; what matters is that the variables are i.i.d. and their variances are proportional to $d^{2}$. This choice affects the following $\widetilde{W}_{2}$-KL correspondence only through an overall constant factor.

Under these preparations, let us recall that KL-divergence between two nearby distributions admits the following expansion:
\begin{eqnarray}
\label{eq:kl-g}
D\left(P_{\xi+\delta\xi}: P_{\xi}\right) = \frac{1}{2}\delta\xi^{\textrm{T}}\Omega\delta\xi + o\left(\left\|\delta\xi\right\|^{2}\right),
\end{eqnarray}
where in the case of Gaussian families with $\xi=\mu$, the r.h.s. is exactly given by up to the quadratic form. Now it is clear that when we take expectation of KL-divergence in Eq.~\eqref{eq:kl-g} w.r.t. $\rho\left(\delta\xi\right)$, its leading order at $d\to 0$ takes
\begin{eqnarray}
\mathbb{E}_{\rho\left(\delta\xi\right)}\left[D\left(P_{\xi+\delta\xi}: P_{\xi}\right)\right] = \frac{a^{2}d^{2}}{24} \textrm{Tr}\left(\Omega\right).
\end{eqnarray}
Meanwhile, the LM-induced transport cost $\widetilde{W}_{2}$ can be obtained in the same way as Gaussian families, namely, 
\begin{eqnarray}
\label{eq:value}
\lim_{d\to 0}\widetilde{W}_{2}^{2} = \frac{d^{2}}{12} \textrm{Tr}\left(\Omega\right).
\end{eqnarray}
Thereby, we obtain a generalized $\widetilde{W}_{2}$-KL correspondence within the leading order at $d\to 0$:
\begin{eqnarray}
\label{eq:klw2}
\widetilde{W}_{2}^{2}\left(P_{\xi},P'\right) = \frac{2}{a^{2}} \mathbb{E}_{\rho\left(\delta\xi\right)} \left[ D\left(P_{\xi+\delta\xi}:P_{\xi}\right)  \right].
\end{eqnarray}


\subsubsection*{Remarks on the Generalization}
The $\widetilde{W}_{2}$-KL correspondence in Eq.~\eqref{eq:rel} provides a transparent physical intuition, i.e., the transport cost between continuous and discrete Gaussian is interpreted as the expected parallel translation of continuous Gaussian, and
in more general statistical submanifolds, the same structure can be interpreted as an information-geometric statement.
The discretization-induced cost measured by the LM-induced transport cost extracts the second-order geometric structure of the KL divergence in terms of Fisher metric at the leading order of $d\to 0$, which is independent of the detailed functional form of the distributions. 
We emphasize that the above leading-order correspondence relies on the assumption that the discretization-induced transport cost is dominated by local fluctuations: This condition can be naturally satisfied for distributions with sufficiently fast-decaying tails (including Gaussian families), where distant contributions to the quadratic transport cost are reasonably suppressed.
In contrast, for distributions exhibiting heavy tails or slow decays, non-local contributions from the tail region may modify the scaling behavior of the $\widetilde{W}_{2}$ in the limit $d\to 0$, potentially requiring corrections including the higher-order terms in $d$.
The present analysis therefore characterizes a universal local correspondence, whose extension to heavy-tailed distributions is left for  our future work.

Through the discussions about Gaussian families and its generalization, we now see that the extension of $\delta\xi$ to random vector is not merely a matter of average, but is structurally necessary.
The trace structure appearing in the $\widetilde{W}_{2}$ expression of Eq.~\eqref{eq:value} can arise from the quadratic form of the KL divergence only if the outer-products of $\delta\xi$ collectively realize an isotropic full-rank structure proportional to the identity matrix, i.e., $\delta\xi\delta\xi^{\mathsf T}\propto d^{2}I_f$. Since the outer product of a single vector is necessarily rank one, it cannot realize this isotropic full-rank structure for systems with $f\ge2$. Therefore, an extension to a probabilistic ensemble of $\delta\xi$ is unavoidable.

We also comment on the role of parametrization and invariance in the present $\widetilde{W}_{2}$-KL correspondence.
While the KL divergence itself is invariant under reparameterizations on the statistical manifold, Eqs.~\eqref{eq:value}-\eqref{eq:klw2} suggest that its expectation under $\rho$ depends on the trace of Fisher metric $\Omega$. 
This does not represent a contradiction, but can be reasonably understood from the following two aspects.
(i) Taking the expectation $\mathbb{E}_{\rho}$ in Eq.~\eqref{eq:klw2} is an \textit{extrinsic} operation living outside the statistical manifold,  where the infinitesimal change in the selected parameter, $\delta \xi$, is linked to the discretization scale $d$ and subsequently extended to a random vector.
(ii) In general, when higher-order terms are present, the KL divergence expanded up to second order in $\delta \xi$, yielding a quadratic form, is explicitly governed by the Fisher metric $\Omega$ associated with the chosen parameter $\xi$. In such cases, once combined with the discretization scale and the extrinsic expectation over parameter variations, the resulting numerical value depends on the chosen parameter and Fisher metric. In the present work, for definiteness, we adopt a discretization based on a uniform hypercubic cell, where the numerical factor $d^{2}/12$ arises from the second moment within a single cell. The $\widetilde{W}_{2}$-KL correspondence in a general discretization scheme is briefly discussed in the Appendix B.

From these viewpoints, the explicit expression $d^{2}\mathrm{Tr}\left(\Omega\right)/12$ generally depends on the chosen parametrization as well as the discretization scheme, yet the correspondence between the LM-induced transport cost and the expected KL divergence can be understood as an invariant structural relation. 
For the Gaussian families considered in the UCN analysis, the Fisher metric is naturally identified through the isomorphism between the parameter $\mu$ on the statistical manifold and the transport coordinate. As a result, the natural parameterization together with the standard quadratic transport cost can lead to $a=1$. While this value of $a$ is not mathematically unique, it corresponds to a canonical choice in which KL divergence on the statistical manifold and the transport cost induce the same local quadratic form, so that no additional rescaling is required. 

Regarding the choice of parametrization, the parameters $\xi$ in the present formulation are taken to be directly identifiable with coordinates on the underlying support (e.g., translational or expectation parameters). This ensures that the Fisher metric $\Omega$ represents the sensitivity of the distribution with respect to physical displacements in the sample space.
While the local quadratic structure of the KL divergence can be expressed for general parametrizations via the pullback Fisher metric, the correspondence with transport-based discretization costs becomes nontrivial when the parametrization is not directly tied to the support coordinates, due to possible degeneracies or distortions in the mapping. For this reason, we restrict the present analysis to parametrizations with a direct physical interpretation in terms of displacements on the support, leaving more general cases for future work.

\subsubsection*{Conceptual Position}
\begin{figure}[h]
\begin{center}
\includegraphics[width=0.99\linewidth]{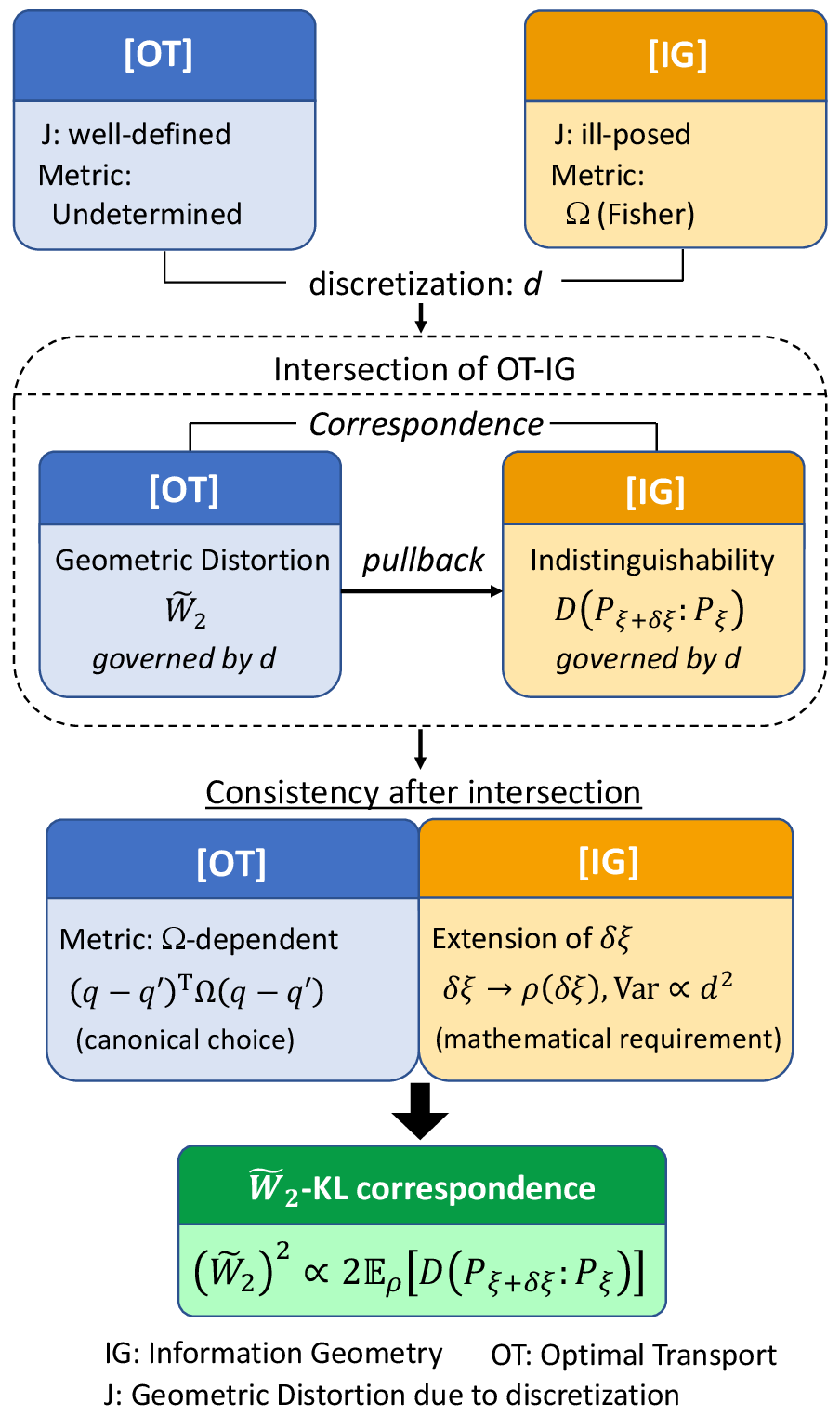}
\caption{Conceptual structure of the transport--information--geometric framework employed in the present work.  }
\label{fig:con}
\end{center}
\end{figure}
In light of the above results, it is instructive to clarify the conceptual position underlying the present framework.
The preceding sections establish that the discretization-induced contribution can be represented locally by an information-theoretic quantity through the derived $\widetilde{W}_{2}$--KL correspondence. This naturally raises a conceptual question: why should an operation performed on the underlying sample space admit an intrinsic description on the statistical manifold?
The key observation is that discretization itself is not an operation defined within information geometry. Rather, it is an external operation acting on the sample space, which inevitably induces a geometric distortion of the underlying probability measure. One possible strategy would be to incorporate such external geometric information directly into information geometry. Such an approach, however, generally depends on how the external geometric structure is introduced, making it difficult to avoid arbitrary choices in the quantification of the discretization-induced geometric distortion.
The present work adopts a different standpoint. Instead of quantifying the discretization-induced distortion directly within information geometry, we first evaluate the geometric effect based on the framework of optimal transport and subsequently investigate how this externally measured quantity \textit{appears} on the statistical manifold.
This viewpoint motivates the following \textit{fundamental postulate}: the geometric distortion induced by discretization should admit a consistent information-theoretic representation. This requirement suggests a bridge between the extrinsic geometry of the sample space and the intrinsic geometry of the statistical manifold.
The key idea is to relate the discretization-induced geometric distortion quantified by transport cost to the indistinguishability—which naturally emerges from the pullback of the LM-induced transport cost onto the statistical manifold—between nearby continuous distributions given by their infinitesimal parameter variations, $D(P_{\xi+\delta\xi}:P_\xi)$.

Under discretization, the lattice scale $d$ and infinitesimal parameter variations $\delta\xi$ capture a common notion of scale dependence.
On the statistical manifold, $\delta\xi$ quantifies the indistinguishability of nearby distributions, while the lattice scale $d$ sets the physically relevant resolution at which such infinitesimal variations can be meaningfully probed.
For the underlying sample space, $d$ simultaneously governs the transport cost measuring geometric distortions due to discretization.
To consistently quantify this scale-dependent correspondence on the transport side, its metric should encode the geometric information that controls infinitesimal indistinguishability on the statistical manifold.
Under this requirement, the admissible transport metrics are no longer arbitrary, but are restricted by the common scale dependence shared with statistical indistinguishability.
Within this restricted class, the cost function is canonically fixed as the Fisher metric up to an overall constant. 
This remaining constant merely sets the unit of the transport cost and can be absorbed into the variance $\left(\propto d^{2}\right)$ of the infinitesimal parameter variations $\rho\left(\delta\xi\right)$, without affecting the correspondence to the local quadratic structure of the KL divergence.
From this viewpoint, $d$ is not merely a discretization parameter, but is explicitly linked to $\delta\xi$. 
Although the link between $d$ and $\delta\xi$, and the subsequent extension of $\delta\xi$ to a random vector can be merged into a single operation, we deliberately separate them to clarify their conceptual roles.

Within the present transport-information-geometric framework, the probability density $\rho\left(\delta\xi\right)$ is intrinsically constrained only through its second moment, whose scaling $\left(\propto d^{2}\right)$ encodes the common discretization scale shared by statistical indistinguishability and transport cost. Under this constraint, the detailed functional form of $\rho\left(\delta\xi\right)$ remains underdetermined. This residual freedom reflects the fact that higher-order structures are not fixed by the intrinsic geometry alone. Instead, they may be selected once additional, extrinsic physical requirements---such as the microscopic origin of the underlying continuous distributions---are specified.

The derived $\widetilde{W}_{2}$--KL correspondence should therefore not be regarded as a coincidence between two unrelated mathematical quantities. 
Rather, it expresses how the geometric cost associated with an external discretization operation appears as a local information-theoretic quantity on the statistical manifold.
 The resultant transport cost naturally provides a characterization of the geometric distortion induced by the discretization, and allows a direct information-geometric interpretation.

A remark on the interpretation of the $\widetilde{W}_2$--KL correspondence is in order: The local quadratic equivalence between the transport cost and the KL divergence follows directly from the adopted construction, where the transport cost is defined via the Fisher metric and the transport plan is specified by the LM scheme. Crucially, the significance of this formulation lies not in the algebraic coincidence of these quadratic forms, but rather in establishing a systematic procedure that quantifies discretization-induced geometric distortion without introducing arbitrary structure. In this sense, the $\widetilde{W}_2$--KL correspondence offers a consistent realization for incorporating extrinsic discretization contributions into an information-geometric description.
The discussed conceptual structure underlying the present transport--information--geometric framework is summarized in Fig.~\ref{fig:con}. The framework provides a route to interpret discretization-induced distortion of local measure through the intersection of optimal transport and information geometry.

\section{Conclusions}
In this work, we investigated the discretization-induced contribution to canonical nonlinearity arising from the representation of continuous Gaussian families on discrete supports, 
based on optimal transport and information geometry.
By focusing on the vanishing discretization limit for the transport cost with its cost function aligned with Fisher metric for Gaussian families, we demonstrated that the resultant LM-induced transport cost $\widetilde{W}_{2}$ is determined solely by the inverse covariance matrix and the discretization scale.

Our central result exhibits that the derived $\widetilde{W}_{2}$ admits a clear information-geometric interpretation: it corresponds to twice the Kullback-Leibler divergence averaged over random parallel translations of the Gaussian induced by the discretization. 
Conceptually, this can provide a precise intuition for what we term unavoidable canonical nonlinearity within the statistical manifold, 
which has been in principle invisible effects solely under the KL-divergence based evaluation on a discrete statistical manifold. 
Finally, the present results indicate that the leading-order correspondence between the discretization-induced transport cost and the expectation of the KL divergence reflects a general transport-information-geometric structure beyond Gaussian families.

\section{Acknowledgement}
This work was supported by JSPS KAKENHI Grant Number 23K04359 and Research Grant from Hitachi Metals$\cdot$Materials Science Foundation.

\section*{Appendix}
\subsection{Numerical verification of the analytic expression for $\widetilde{W}_{2}$ at finite $d$}
\begin{figure}[h]
\begin{center}
\includegraphics[width=0.98\linewidth]{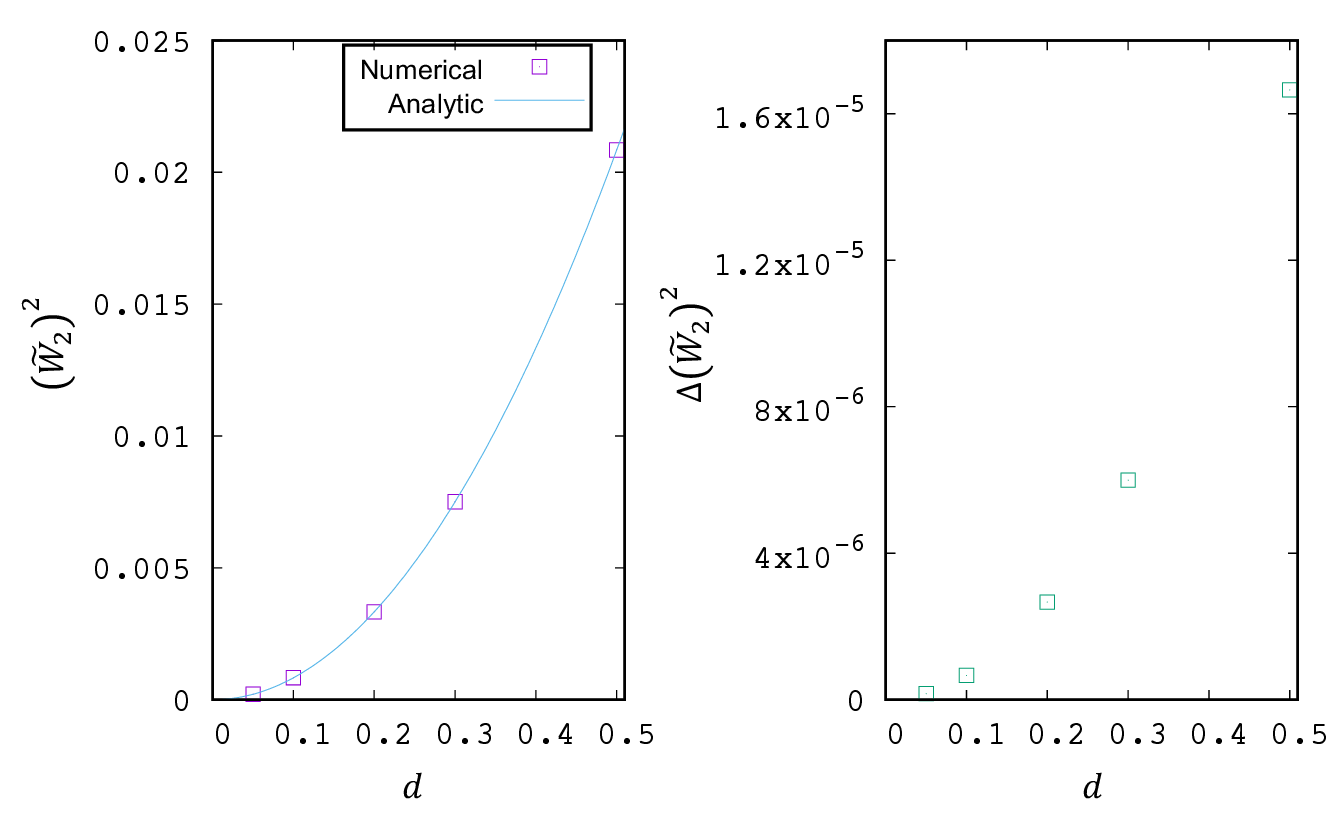}
\caption{
Left: $d$-dependence of the squared LM-induced transport cost $\widetilde{W}_{2}^{2}$ obtained from the analytic expression Eq.~\eqref{eq:w2} (``Analytic'') and from the numerical evaluation of Eq.~\eqref{eq:w-int} (``Numerical''). 
Right: $d$-dependence of the difference between the numerical and analytic values of $\widetilde{W}_{2}^{2}$, defined as the numerical value minus the analytic one.
}
\label{fig:w2}
\end{center}
\end{figure}
In order to briefly confirm the accuracy of the derived transport cost, we compare the analytical expression given in Eq.~\eqref{eq:w2} with the original integral expression in Eq.~\eqref{eq:w-int} for a simple one-dimensional ($f=1$) Gaussian case.
We set the standard deviation to $\sigma = 1$. 
The numerical evaluation of Eq.~\eqref{eq:w-int} is performed using the trapezoidal rule with uniform discretization. 
The integration range is restricted to $q \in \left[-5\sigma, 5\sigma\right]$, and the representative point $q'_k$ of each discretization cell is chosen as its midpoint.
The numerical results shown in Fig.~\ref{fig:w2} indicate that the derived analytic expression for $\widetilde{W}_{2}$ accurately reproduces the original integral form, particularly in the small-$d$ regime.

\subsection{$\widetilde{W}_{2}$-KL Correspondence for a General Discretization Scheme}
Here, we briefly discuss the $\widetilde{W}_{2}$-KL correspondence for a general discretization scheme.
Let $\omega \subset \mathbb{R}^{f}$ be an arbitrary bounded convex set with volume $V_{\omega}$.
We consider a discretization of $\mathbb{R}^{f}$ by translating $\omega$, and take the limit $V_{\omega} \to 0$, in which the discretization scale vanishes.
Without loss of generality, the coordinate $u$ is chosen such that its mean over $\omega$ vanishes.
We define the second-moment matrix of $\omega$ as
\begin{eqnarray}
M = \frac{1}{V_{\omega}} \int_{\omega} uu^{\mathrm{T}} du.
\end{eqnarray}
Then, to leading order in the discretization scale, the LM-induced transport cost is given by
\begin{eqnarray}
\widetilde{W}_{2}^{2}
=
\frac{1}{V_{\omega}}
\int_{\omega}
u^{\mathrm T}\Omega u\,du,
\end{eqnarray}
where $\Omega$ denotes the quadratic transport cost aligned with the Fisher metric.
Using the standard trace identity
\begin{eqnarray}
u^{\mathrm T}\Omega u
=
\mathrm{Tr}
\left(
\Omega uu^{\mathrm T}
\right),
\end{eqnarray}
we obtain
\begin{eqnarray}
\widetilde{W}_{2}^{2}
&=&
\frac{1}{V_{\omega}}
\mathrm{Tr}
\left(
\Omega
\int_{\omega}
uu^{\mathrm T}du
\right)
\nonumber\\
&=&
\mathrm{Tr}
\left(
\Omega M
\right).
\label{eq:w2-1}
\end{eqnarray}

On the other hand, let $\rho\left(\delta\xi\right)$ denote the probability distribution of infinitesimal parameter variations, and let its covariance matrix be
\begin{eqnarray}
\Lambda = \mathbb{E}_{\rho}\left[\delta\xi \delta\xi^{\mathrm{T}}\right].
\end{eqnarray}
Then the expectation of the quadratic form for the KL divergence expansion reads
\begin{eqnarray}
\label{eq:kl-1}
2\mathbb{E}_{\rho}\left[ D\left(P_{\xi+\delta\xi}:P_{\xi}\right) \right]
= \mathbb{E}_{\rho}\left[\delta\xi^{\mathrm{T}}\Omega\delta\xi\right]
= \mathrm{Tr}\left(\Omega\Lambda\right).
\end{eqnarray}
Comparing Eqs.~\eqref{eq:w2-1} and~\eqref{eq:kl-1}, we see that the $\widetilde{W}_{2}$-KL correspondence holds whenever
\begin{eqnarray}
\mathrm{Tr}\left(\Omega\Lambda\right)
\propto
\mathrm{Tr}\left(\Omega M\right),
\end{eqnarray}
where the proportionality constant is a scalar independent of the quadratic form defined by $\Omega$.
In particular, a sufficient condition for this proportionality is
\begin{eqnarray}
\Lambda \propto M.
\end{eqnarray}
As a special case, for a hypercubic discretization of side length $d$,
\begin{eqnarray}
M = \frac{d^{2}}{12} I,
\end{eqnarray}
which leads to the consistent result of Eq.~\eqref{eq:w2}.

Therefore, in a general discretization scheme, the probability distribution $\rho\left(\delta\xi\right)$ should be chosen so that its covariance reflects the geometry of the discretization cell through its second-moment structure $M$.

\end{document}